\documentclass[lettersize,journal]{IEEEtran}
\usepackage{amsmath,amsfonts}
\usepackage{algorithmic}
\usepackage{algorithm}
\usepackage{array}
\usepackage[caption=false,font=normalsize,labelfont=sf,textfont=sf]{subfig}
\usepackage{textcomp}
\usepackage{stfloats}
\usepackage{url}
\usepackage{verbatim}
\usepackage{graphicx}
\usepackage{cite}
\usepackage{eso-pic}
\begin{document}
\AddToShipoutPictureFG*{
\AtPageUpperLeft{
\hspace*{0.05\paperwidth}
\raisebox{-1.5cm}{
\parbox{0.9\paperwidth}{
\centering
\footnotesize
This is the author's version of a manuscript accepted for publication in \textit{IEEE Transactions on Plasma Science}. 
The final version will be available via IEEE Xplore.
}
}
}
}
\title{Experimental Results from Early Non-Planar NI-HTS Magnet Prototypes for the Columbia Stellarator eXperiment (CSX)}

\author{D. Schmeling\textsuperscript{1}, M. Russo\textsuperscript{1}, B. T. Gebreamlak\textsuperscript{1}, T. J. Kiker\textsuperscript{1}, A. R. Skrypek\textsuperscript{1}, A. R. Hightower\textsuperscript{1}, J. Xue\textsuperscript{1}, S. Chen\textsuperscript{1}, S. Sohaib\textsuperscript{1}, C. Martinez\textsuperscript{1}, K. F. Richardson\textsuperscript{1}, L. Filor\textsuperscript{1}, S. Komatsu\textsuperscript{1}, L. Liu\textsuperscript{2}, C. Paz-Soldan\textsuperscript{1}
%IEEE Publication Technology,~\IEEEmembership{Staff,~IEEE,}
        % <-this % stops a space
%\thanks{This paper was produced by the IEEE Publication Technology Group. They are in Piscataway, NJ.}% <-this % stops a space
\thanks{\textsuperscript{1}Columbia University, New York, NY, USA.}%
\thanks{\textsuperscript{2}Pomona College, Claremont, CA, USA.}%
\thanks{Manuscript received xxxx; revised xxxx.}}

\markboth{IEEE Transactions on Plasma Science}%
{Schmeling \MakeLowercase{\textit{et al.}}: Experimental Results from Early Non-Planar NI-HTS Magnet Prototypes for CSX}


\maketitle

\begin{abstract}
The Columbia Stellarator eXperiment (CSX) is an upgrade of the Columbia Non-neutral Torus (CNT) that aims to demonstrate a university-scale, quasi-axisymmetric stellarator using high-temperature superconducting (HTS) technology at an on-axis magnetic field target of 0.5 T. Due to the strain sensitivity of ReBCO, adapting it to non-planar geometries requires new winding, structural, and cooling strategies. We report on the results of a staged prototype program (P1, P2, P3) employing 3D-printed, sectional aluminum coil frames with winding channels, gimballed constant-tension winding mechanics, and solder potting for radial current redistribution and passive quench mitigation. The first prototype, P1 (planar elliptical, double‑pancake) tested additive manufacture, sectional joining and baseline winding, achieving predicted fields at 77 K. P2 (non‑planar, higher strain) was wound to 42 turns, energized at 30–40 K to produce expected magnetic fields, and studied thermal gradients and resistance at up to 110 A (4.5 kAt). Design evolution in P3 introduces concave geometry with dual double-pancakes and 200 turns, and has been commissioned at 20 K, with high-field characterization ongoing. In parallel, sub‑$\mu \Omega$ lap joints have been developed. Together, these results de‑risk manufacturing, cooling interfaces, quench management, and diagnostics, paving the way for full‑size non‑planar HTS stellarator coils for CSX.

\end{abstract}


\section{Introduction}
\IEEEPARstart{T}{he} widespread adoption of high-temperature superconductors (HTS) has contributed to a surge in fusion companies aiming to achieve sustainable burning plasmas through magnetic confinement, enabled by the use of HTS-based high-field magnets \cite{hartwig, nash}. However, the HTS tape form factor makes it better suited to planar magnets, and the adaptation to non-planar stellarator magnets presents several challenges. We present work being done to design and test no-insulation no-twist (NINT) non-planar HTS magnets for the Columbia Stellarator eXperiment (CSX), a quasisymmetric device under development composed of two optimized non-planar interlocked HTS magnets and two copper poloidal field magnets \cite{baillod}. This is an upgrade of an existing experiment, the Columbia Non-neutral Torus (CNT), aiming for a quasi-axisymmetric plasma with a higher field of $0.5$ T on-axis and $5.3$ T peak on-coil field. It will build on existing CNT infrastructure, including the vacuum vessel and components, poloidal field magnets, RF heating system, and diagnostics \cite{pedersen}. 

Prototype and eventual full-scale NINT magnets consist of HTS tape (YBCO) directly wound in non-planar channels on 3D-printed aluminum frames, followed by vacuum solder-potting. A gimballed winding mechanism maintains constant tension and mitigates strain during the winding process. The NINT magnet fabrication process stands in contrast to the approaches taken by other groups designing non-planar magnets. For instance, Type One Energy, in collaboration with MIT, is pursuing a twisted cable-in-conduit conductor (CICC) approach, which is possible due to their larger bend radii \cite{riva}. While more technically challenging, the NINT technique allows for smaller bend radii and higher current densities, ideal for university-scale devices \cite{huslage3}.

\begin{figure}[t!]
\centering
\includegraphics[width=3.5in]{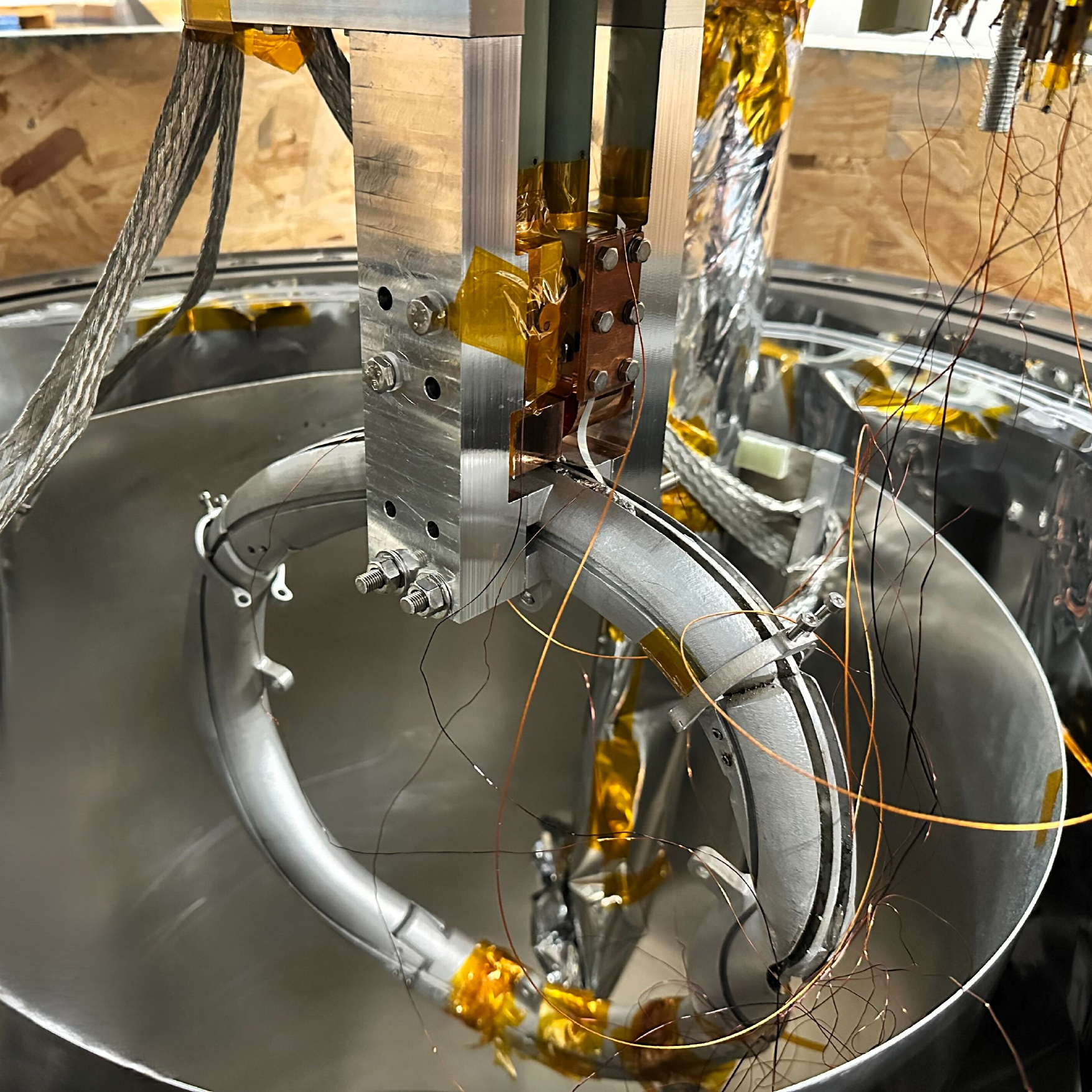}
\caption{Prototype 2 (P2) HTS coil installed inside the cryogenic test-stand. Voltage taps and silicon diodes for diagnostics are visible. The HTS strands interface with copper leads above the coil, which connect to HTS-110 leads.}
\label{fig_1}
\end{figure}

At present, prototype non-planar HTS magnets have been successfully tested in liquid nitrogen, with ongoing efforts to achieve higher on-axis fields of $0.5$ T in a cryogenic test-stand. The prototypes are conductively cooled to $20$ K target temperatures using a cold head, with numerous diagnostics and control systems in place. Because low-resistance HTS tape lap joints are required for the final magnets, the joints are produced and tested in liquid nitrogen.%. Diagnostics and controls include a high-current power supply, scanning Hall probes, temperature sensors, and ion pressure gauges. Due to the requirement of low-resistance HTS tape lap joints in the final magnets, joints are optimized and tested in liquid nitrogen, with future tests planned in the cryogenic test-stand at $20$ K.

The paper presents the design and details of current prototypes in Section II, followed by the design of the experimental test-stand for cryogenic testing in Section III. Section IV will briefly describe the joint testing methodology and results. Section V will present results from the testing of the prototype magnets. Finally, Section VI will summarize the testing campaign for these prototypes and look towards future designs. This work aims to enable the construction of a strain- and field-optimized university-scale HTS stellarator, addressing critical engineering challenges in the adaptation of HTS technology for stellarator configurations.

\section{Magnet Design and Enabling Technologies}

The prototype and planned full-scale magnets are designed from 3D-printed aluminum alloy (AlSi10Mg) bobbins with HTS wound in channels under tension before being solder potted. Three prototypes of increasing complexity have been sequentially designed to test and de-risk steps in the production of the final magnets, henceforth referred to as P1, P2, and P3. %and are undergoing testing in different conditions. These are henceforth referred to as P1, P2, and P3. 

Rare-earth Barium Copper Oxides (ReBCO) are a widely used HTS material, but suffer from significant performance degradation under excess strain, including torsion and hard-way bending. The development of strain-optimization to reduce strain in HTS tapes has been foundational to allowing the application of HTS to non-planar NINT magnets, and thus for the development of stellarator configurations \cite{paz-soldan,huslage1,huslage2}. 

\begin{figure}[!hb]
\centering
\includegraphics[width=3.5in]{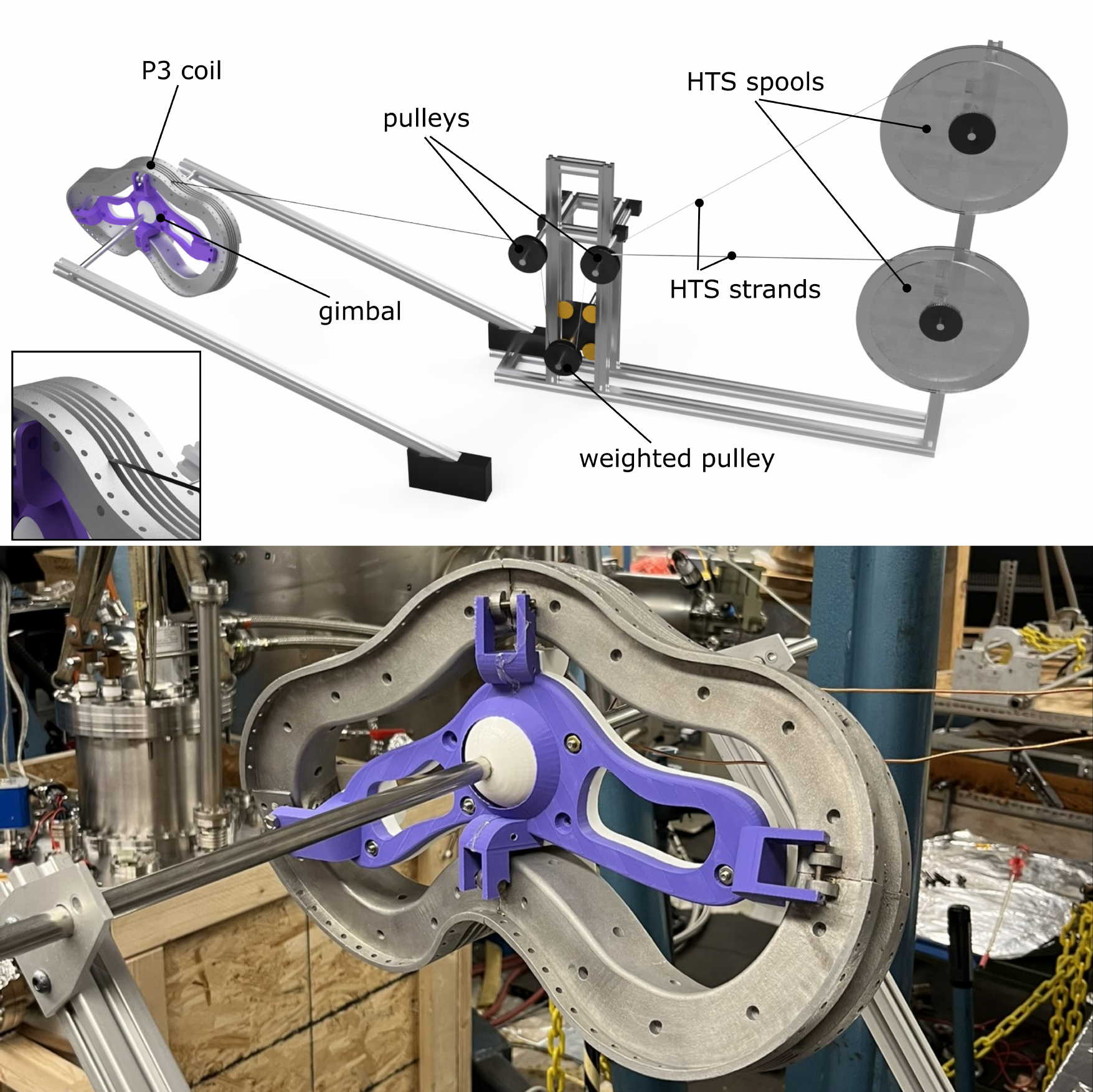}
\caption{CAD rendering of winding rig (top) and photograph  of P3 prototype (bottom) mounted on the gimbal. A ball joint allows variation of winding angle as the coil is rotated, to enable HTS placement in 3D-printed channels. The use of one or both spools is possible when winding.}
\label{gimbal}
\end{figure}

Winding such non-planar NINT magnets presents a new challenge, as the tape must be wound into the channels without bending the tape excessively. A gimballed winding system was thus developed, as shown in  Fig.~\ref{gimbal}. It allows the magnet to easily change winding angles as well as slide along its axis, ensuring the tape is always perpendicular to the channel to  minimize strain. Solder paste is dispensed onto the tape surface using a syringe and spread evenly across the surface by applying pressure. This system works well for magnets with no or minimal concavity (P1, P2), where controlled constant tension can be applied to ensure the tape stays in the channel. For magnets with concave sections, as a number of candidate final CSX magnets have, novel methods need to be used. These include systems to hold the tape down in the concave sections, winding under compression in those areas. The viability of such systems is currently under investigation in P3.

\begin{figure}[h]
\centering
\includegraphics[width=3.5in]{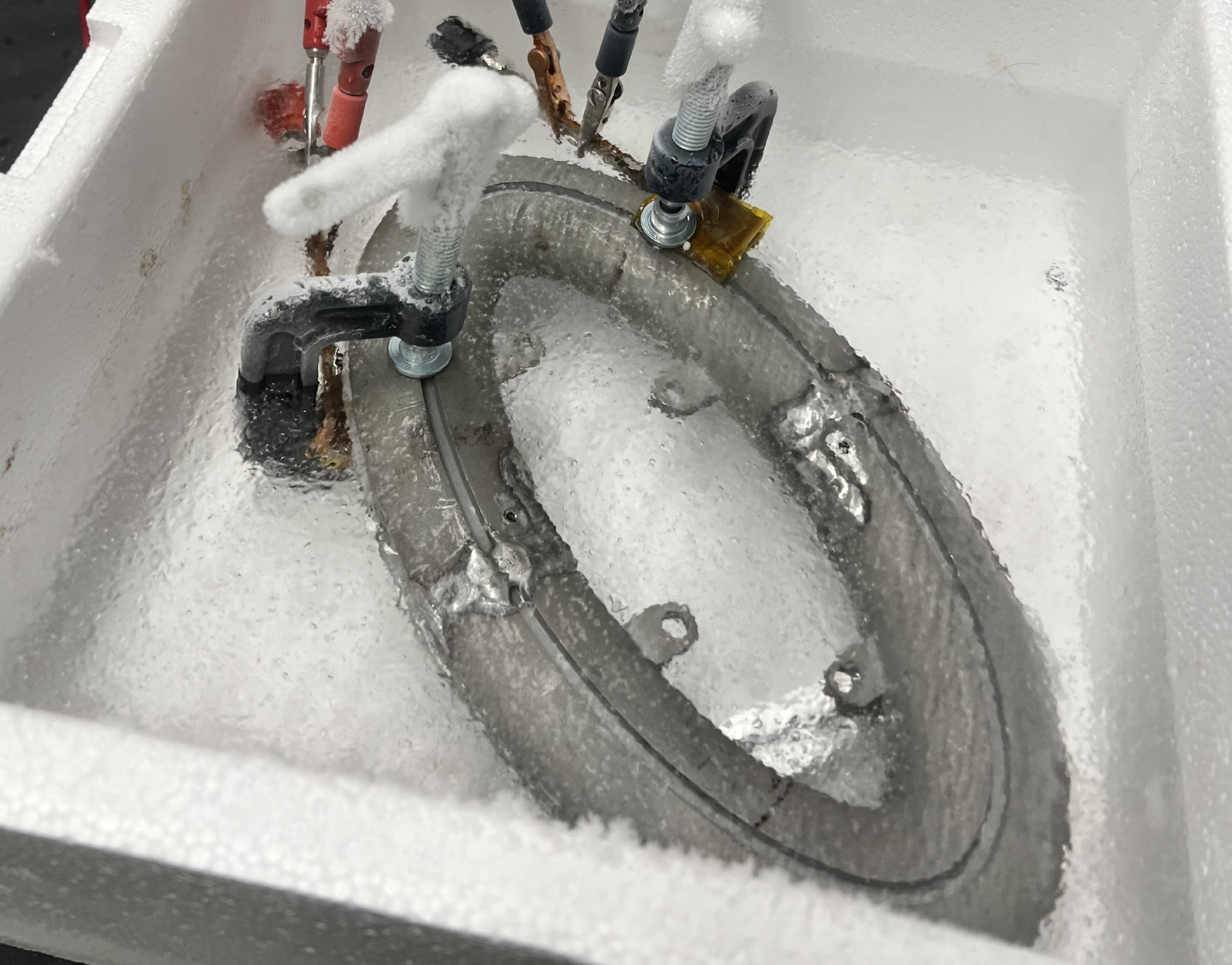}
\caption{77 K tests of elliptical prototype magnet P1, shown immersed in a liquid nitrogen bath during characterization. The coil consists of 22 turns of HTS tape, with copper leads attached for resistance measurements.}
\label{fig_2}
\end{figure}

Furthermore, the quench behavior of HTS means magnets that transition out of superconductivity (due to local defects, hot-spots, or exceeding the critical current of the tape) can undergo rapid heating, posing the risk of further damage to the magnet. One method of passive quench mitigation involves solder potting, which allows current to dissipate radially through the tape-stack and into the bobbin at local hot-spots. This redistribution of current and heat reduces the risk of damaging the magnet. Consequently, all magnet designs use solder potting to bond the HTS layers into stacks. In this process, ChipQuik® Sn63Pb37 solder paste is dispensed onto the tape during winding and then reflowed at $190~^\circ$C.

The manufacture of the 3D-printed frames was first investigated by P1, which is a small-scale, planar elliptical magnet, with two channels for double-pancake winding. Due to the small print-bed size of commercially available 3D printing/additive manufacturing providers, none of the magnets could be printed as a single piece. P1 was printed in two sections, with dovetail joints between the sections to allow for mechanical coupling. The sections were then welded together for added support. P1 was wound with $22$ turns ($11$ per channel), with ChipQuik® Sn42/Bi57.6/Ag0.4 solder paste spread between layers before baking at $190~^\circ$C. %It was then tested in liquid nitrogen, yielding magnetic fields consistent with predictions.

\begin{figure}[hb!]
\centering
\includegraphics[width=3.5in]{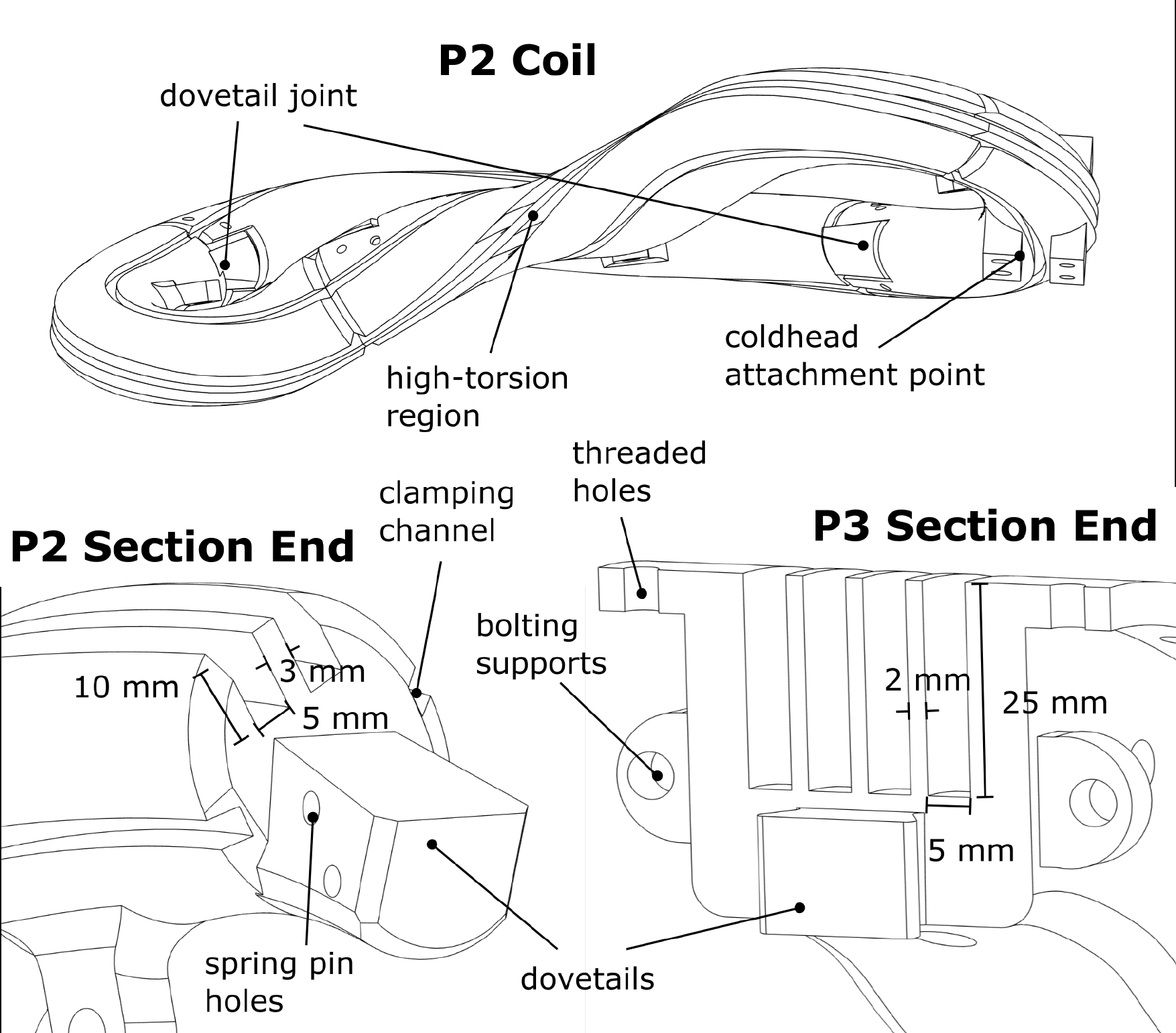}
\caption{Illustrations of prototype cross-sections highlighting HTS channel geometry and dovetail joint evolution from P2 to P3.}
%Illustrations of key design points for prototype magnets. The difference in cross-section from 2- to 4-channel design can be seen between P2 and P3, as well the change in dovetail design.}
\label{cross_section}
\end{figure}

P2 was the next prototype, designed to test winding of non-planar, high-strain coils. It was printed in four sections with two channels for the double-pancake, with the dovetail joints held together using spring pins instead of welding. The spring pin method, and subsequent bolting method on P3, were used due to a lack of in-house welding capabilities. Bolting provides improved modularity and repeatability during  prototyping, and allows controlled preload across  interfaces to ensure adequate thermal contact while preserving geometric tolerances. A number of design features are shown in Fig.~\ref{cross_section} for both P2 and P3 (P1 is not shown as it shares the same cross-section as P2). It was wound with $42$ turns out of the expected $100$ as a high-torsion region prevented winding of additional HTS layers due to excessive tape rotation in the channel. Solder potting proceeded similarly to P1, using ChipQuik® Sn63Pb37 due to the pre-applied solder coating on the tape, and baked at $190~^\circ$C for $30$ minutes. Liquid nitrogen tests confirmed superconducting operation prior to $20$ K testing.

As a result of the challenges faced in the high-torsion area of P2, and the potential concave sections of the final CSX magnets, P3 was designed to reduce local torsional strain by redistributing curvature into concave coil sections. This allows the testing of two high-risk aspects of the magnets, namely the winding of concave sections and management of high-torsion sections.
Additionally, in order to achieve the higher fields of $0.5$ T on-axis, the final magnets are being designed with two double-pancakes, enabling more tape to be wound on the magnet without sacrificing coil-to-plasma distance. P3 thus has two double-pancakes with channels of similar depth to the final CSX magnets, to study magnets with such cross-sections. The channels are $25$ mm deep, $5$ mm wide, with $2$ mm separators. P3 is designed to be a high-field magnet, aiming to achieve fields on the order of $0.5$ T on-axis and $2$ T on the coil with $200$ turns of HTS. In addition to dovetail joints, P3 features a design allowing sections to be bolted together with indium for an improved thermal interface. Finally, to further reduce the risk of quenching, P3 used parallel-wound non-insulated (PWNI) HTS architecture, allowing improved current sharing between co-wound HTS tapes in the same channel \cite{rogers,kobayashi,xue}.

\section{Testing Infrastructure and Diagnostics}

%The testing of magnet prototypes is central to the experimental side of CSX, to ensure the success of final magnets and to identify and overcome challenges. As a result, a specialized test-stand and experimental protocols have been developed to test the prototype magnets beyond verification of windability.

\begin{figure}[h!]
\centering
\includegraphics[width=3.5in]{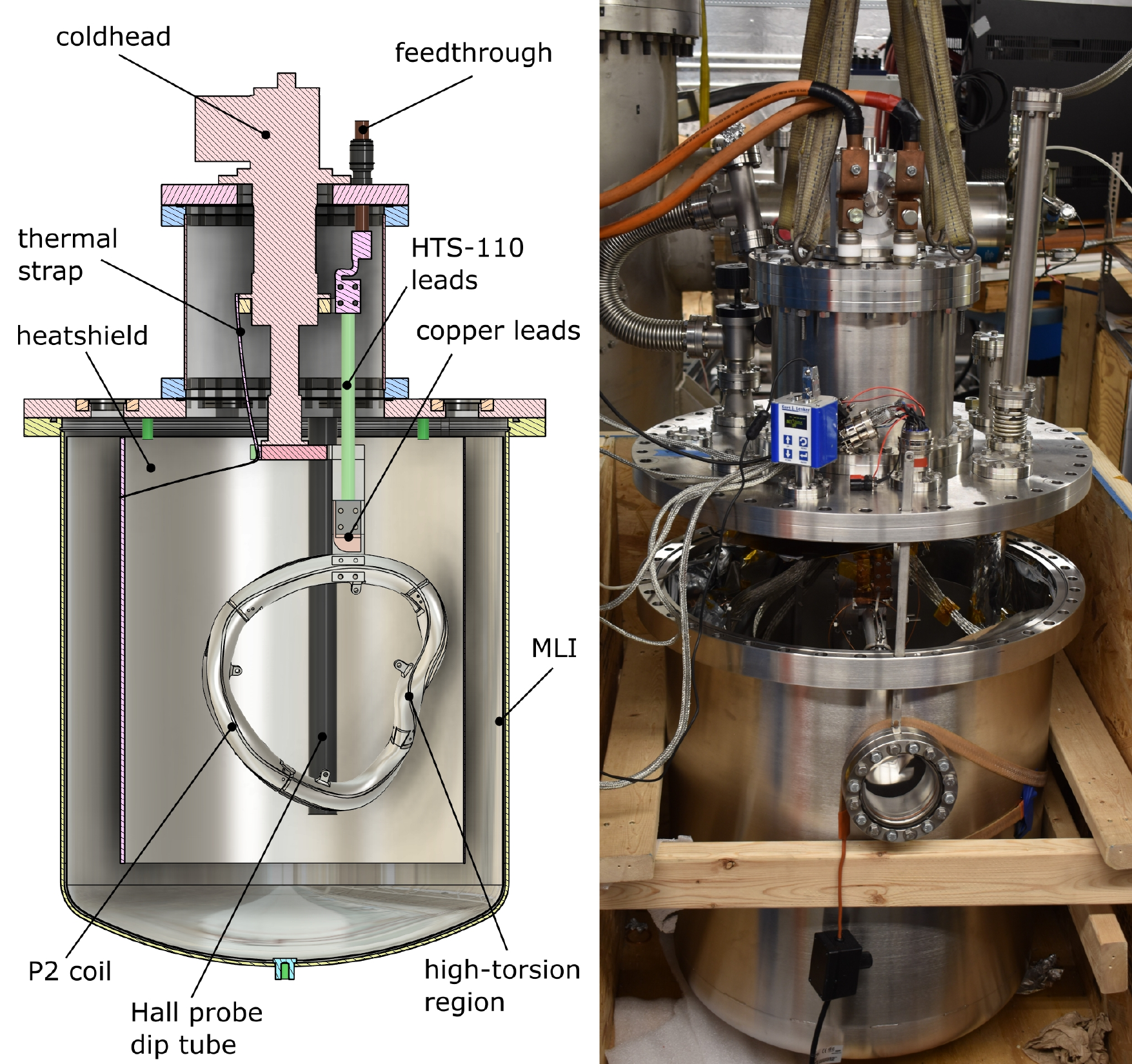}
\caption{CAD cross-section (left) and photograph (right) of the cryogenic test-stand with P2 installed. Rapid installation and testing is enabled by the modular design. Feedthroughs on the top flange accommodate high-current leads, vacuum sensors, and other diagnostics.}
\label{test_stand}
\end{figure}

A modular 20 K test-stand was developed to evaluate prototypes beyond 77 K verification. Preliminary tests are conducted by submerging the entire magnet in liquid nitrogen, as shown for P1 in Fig.~\ref{fig_2}, and making magnetic field and resistance measurements. Subsequently, the magnets are installed in an experimental $20$ K test-stand, where they interface with a Sumitomo 408S cold head. P2 can be seen installed prior to testing in Fig.~\ref{fig_1}, and the experimental layout is illustrated in Fig.~\ref{test_stand}. Current feedthroughs, rated for $800$ A, provide current from a Lambda TDK high-current power supply capable of $1$ kA at $10$ V. 

In the chamber, the current is transferred from the feedthroughs to HTS-110 G10-insulated leads, which are conductively cooled through a sapphire interface attached to the first stage of the cold head. The HTS-110 leads make contact with copper leads attached to the second stage using another sapphire interface. This allows the transition from first stage temperatures of around $50$ K to second stage temperatures of $20$ K, ensuring the HTS leads are superconducting while maintaining electrical insulation. The copper leads then interface with bare $4$ mm HTS from the magnet prototype, which is clamped to the copper with indium foil for improved thermal and electrical contact. %cooling.

To ensure adequate thermal insulation from the outside environment, a rough vacuum of $10^{-3}$ Torr is achieved using a roughing pump. Subsequent cryopumping from the cold head brings the final pressure to below $10^{-6}$ Torr. Two thermal radiation shields are used to further isolate the system. The first is composed of $6$ layers of multilayer insulation (MLI) manufactured in-house. The MLI layer provides decent coverage of the solid angle of the magnet. The second, inner layer, providing nearly full solid angle coverage, is an actively cooled aluminum heat-shield, composed of a cylinder with top and bottom made from $1$ mm thick aluminum sheet. All sections of the heat-shield are connected to the first stage of the cold head using copper braid, and a final temperature of around $100$ K is achieved on the heat-shield.

The diagnostics currently installed in the test-stand include five Scientific Instruments SI-540 silicon diodes for temperature sensing, a PCE-MFM 3000 gaussmeter, voltage taps soldered to the HTS of the magnet, and pressure gauges. Of the $10$ ports on the top flange, $4$ are used for diagnostics and $2$ for vacuum control and management. The gaussmeter is not vacuum compatible and interfaces with the test-stand through a dip tube at atmosphere which can be placed at variable radii and angles from the magnet axis. The dip tube was installed on the port allowing for the minimum radial distance, placing the Hall probe along the magnet axis at a distance of $0.19$ m. Planned additions include an RGA, and additional Hall sensors and voltage taps. All diagnostics are interfaced through LabVIEW, with voltage taps for resistance measurements interfaced with a Keithley DMM6500 multimeter.

\section{Joint Testing Methodology}

For the planned fields of around $0.5$ T on-axis for the final CSX magnets, tape requirement estimates are on the range of several kilometers, depending on operating temperature, current, and margin of safety. Consequently, low-resistance HTS lap joints between sections of tape are essential, and the design of effective joints is being optimized. Joints are created by overlapping strands of HTS and soldering these together using SnPb solder \cite{lalitha, chen}. Due to the SnPb pre-coating on the HTS, low-temperature solder is not recommended. 

\begin{figure}[hb!]
\centering
\includegraphics[width=3.5in]{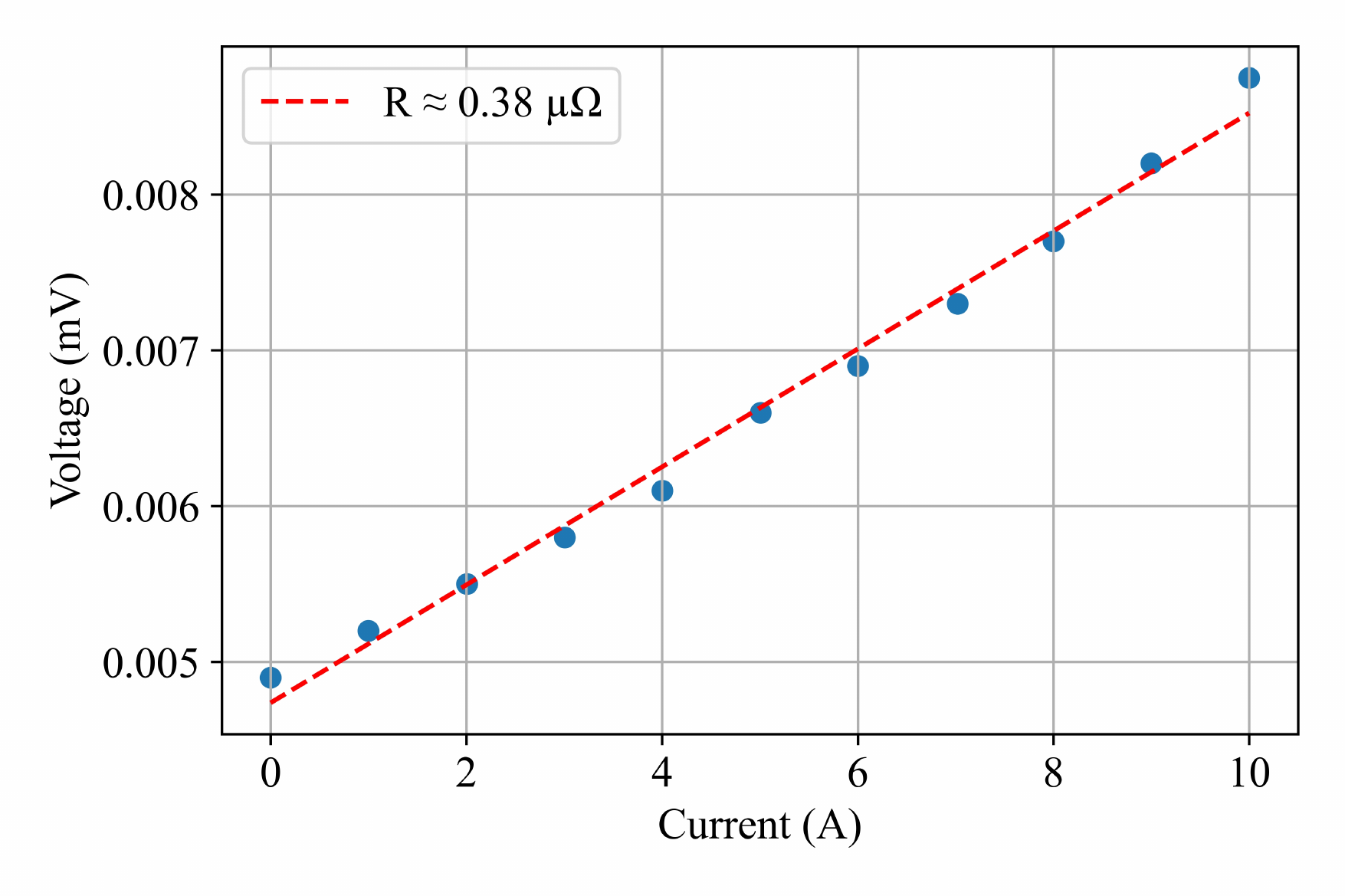}
\caption{Resistance characterization of a $30$ mm HTS lap joint measured at $77$ K. The slope of the V–I curve indicates a joint resistance of approximately $380$ n$\Omega$, with good linearity across the $0–10$ A range.}
\label{joints}
\end{figure}

The soldering process involves overlaying selected lengths of HTS with ChipQuik® Sn63Pb37 solder paste between the layers, and soldering these together under compression using a $5.5$ cm rectangle of aluminum heated to $190~^\circ$C using a soldering iron. The alignment of the two tape layers is assisted using a G10 block with a $4$ mm wide groove. The joints are then tested using a four-point probe setup in liquid nitrogen, with joints of $30$ mm yielding a resistance of $380$ n$\Omega$ (Fig.~\ref{joints}).
Preliminary tests of a resistance–length relation show a rapid initial decline followed by a plateau, allowing an optimal joint length to be found.

In addition to tests at $77$ K, a setup to test joints at $20$ K in the cryogenic test-stand is being designed. This setup will allow tests at higher critical currents, improving joint characterization in conditions relevant to the final magnets. %and the implementation of additional variables such as strain (bending, twisting) and background magnetic fields. This enables improved joint characterization in conditions relevant to the final magnets.

\section{Experimental Results }

%The prototypes P1 and P2 have been wound and tested by the time of writing. Early testing of P3 at $20$ K has confirmed its operatio, with results from high-field tests to be reported in a later paper. Following the winding of P1 and P2, each was first tested at $77$ K, measuring resistance using the four-point probe method, and supplying up to $12$ A to measure the on-axis field. P1 had a total resistance of $30.1~\mu \Omega$ and a field of $0.55$ mT at $10$ A, meeting expected values. %P2 yielded a similar resistance of $33.1~\mu \Omega$, with an on-axis field of $2.1$ mT at $12$ A. The field measurements were in the expected ranges for both prototypes compared to calculated fields.

The three prototypes P1, P2 and P3 have been wound at the time of writing, with P1 and P2 having undergone rigorous testing. Early testing of P3 at $20$ K has confirmed its operation, with results from high-field tests to be reported in a later paper. Following the winding of P1 and P2, each was first tested at $77$ K, measuring resistance and on-axis magnetic field at up to $12$ A. P1 had a total resistance of $30.1~\mu \Omega$ and a field of $0.55$ mT at $10$ A, meeting expected values.

\begin{figure}[hb!]
\centering
\includegraphics[width=3.5in]{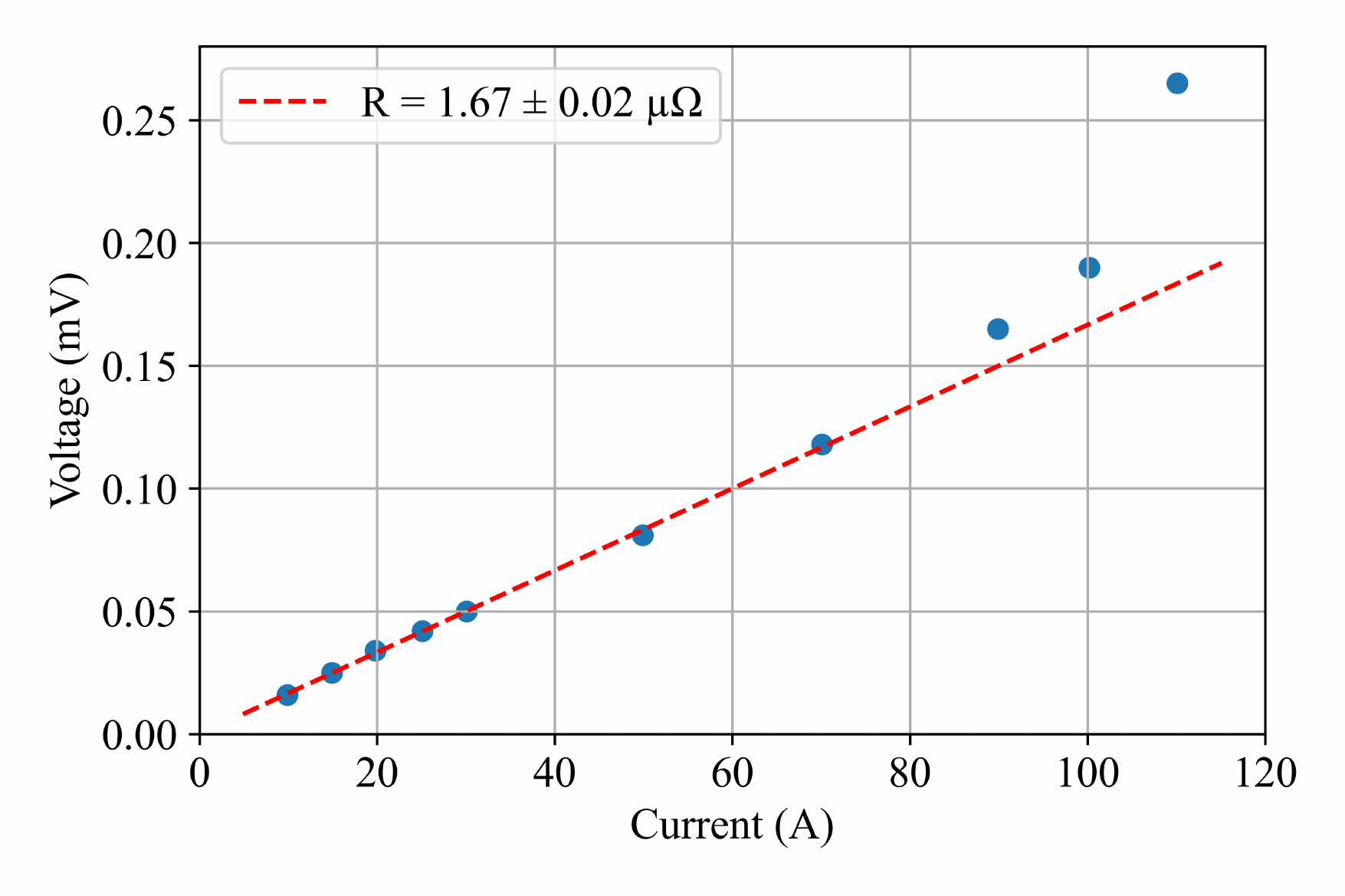}
\caption{Resistance of P2 measured using voltage taps soldered to the HTS tape near the copper leads. A linear fit up to $70$ A indicates a resistance of $R = (1.67 \pm 0.02)~\mu \Omega$, with an approach to quench evident at over $100$ A.}
\label{p2_res}
\end{figure}
\begin{figure}[hb!]
\centering
\includegraphics[width=3.5in]{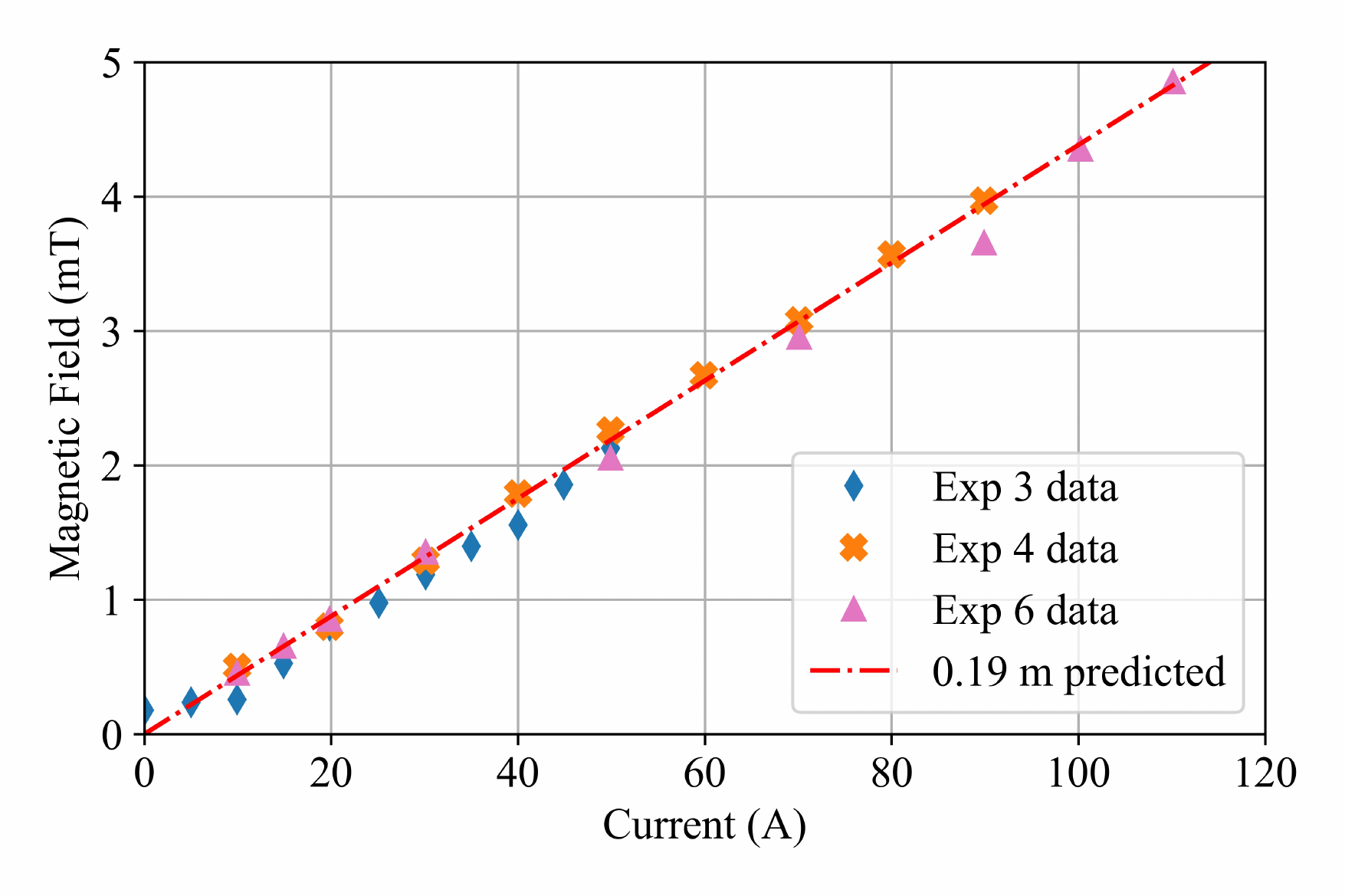}
\caption{P2 magnetic field measured at a distance of 0.19 m from coil center across multiple experimental runs, compared with Biot–Savart predictions.}
\label{p2_field}
\end{figure}
Following basic testing at $77$ K, P2 was installed and tested in the $20$ K cryogenic test-stand. Temperature sensors were placed at the top and bottom of the magnet, as well as the magnet HTS-copper interface, where resistive heating was expected. These locations achieved final temperatures of $32$ K, $42$ K, and $48$ K, respectively. Voltage taps soldered close to the HTS leads of the magnet were used to measure the voltage corresponding to different currents shown in Fig.~\ref{p2_res}, yielding a resistance of $R = (1.67 \pm 0.02)~\mu \Omega$. This is the resistance across the entire magnet, including soldered HTS joints but excluding copper leads. As shown in Fig.~\ref{p2_field}, the magnet was operated across multiple runs at up to $110$ A, with a magnetic field of $4.5$ mT. Attempts to reach $120$ A resulted in gradual magnet quenching, as can be seen in the pronounced increase in resistance at high current.

\begin{figure}[h!]
\centering
\includegraphics[width=3.5in]{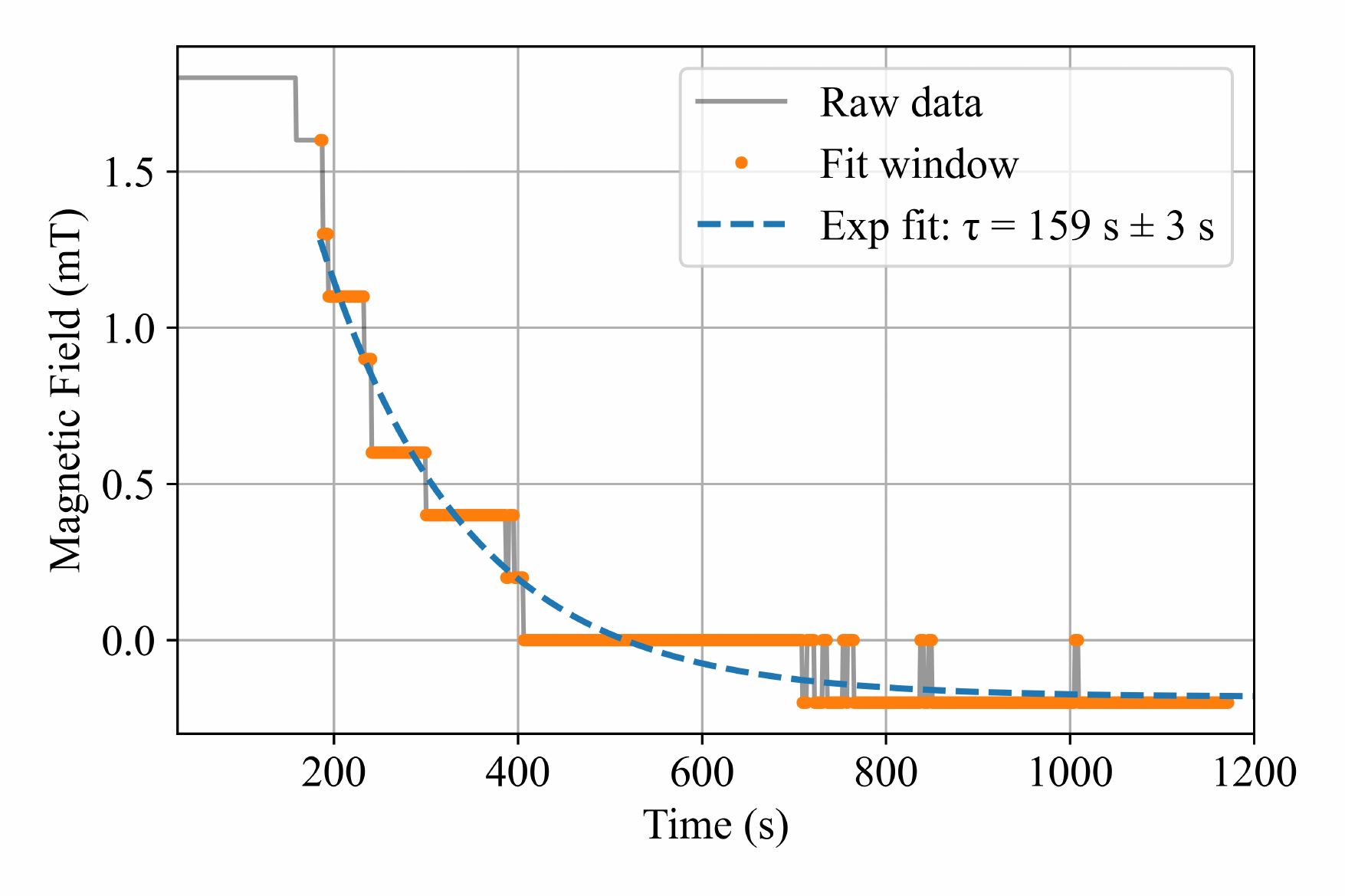}
\caption{Exponential decay of magnetic field following current shut-off, measured using a Hall probe. A fit to the field decay yields an $L/R$ time constant of $\tau = (159 \pm 3)$ s.}
\label{p2_mag}
\end{figure}
\begin{figure}[h!]
\centering
\includegraphics[width=3.5in]{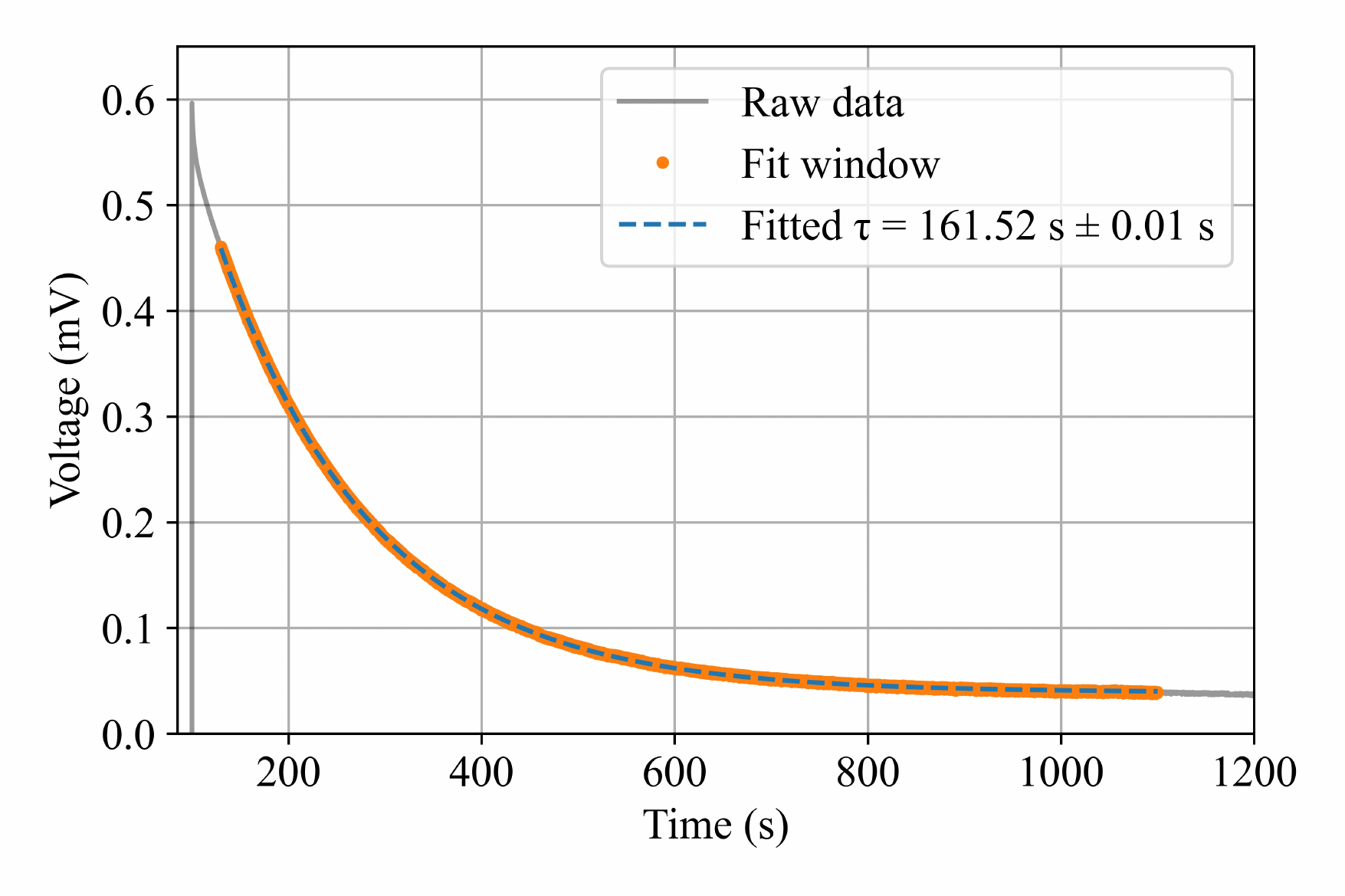}
\caption{Voltage decay across the magnet leads following current shut-off. Fitting yields an $L/R$ time constant of $\tau = (161.52 \pm 0.01)$ s.}
\label{p2_volt}
\end{figure}
Temperatures across the magnet increased as current was supplied, with the HTS-copper interface and other higher-resistance sections having the greatest increase in temperature. Due to excessive heating on one of the clamped copper-HTS interfaces during early testing, the connection was broken as the HTS burnt through. Repairs were implemented using improved clamping and thermal management. However, due to the addition of soldered HTS leads in this region, it is possible these limited the HTS critical current due to resistive heating.

By rapidly shutting off the current supply, the corresponding decay of the magnetic field (Fig.~\ref{p2_mag}) and the voltage across the leads (Fig.~\ref{p2_volt}) was measured to determine the magnet's $L/R$ time $\tau$. This was achieved by fitting an exponential to the decay, which yields $\tau = (161.52\pm 0.01)$ s according to %. This was found to be approximately $2.7$ minutes.
\begin{equation}
V(t) = V_0e^{-\frac{t}{\tau}}.
\end{equation}
The magnetic field yielded $\tau = (159 \pm 3)$ s, in close agreement with the voltage-based $L/R$ measurement despite the limited resolution.

Finally, by ramping the current in the magnet at a constant rate and measuring the back emf, the inductance could be calculated using 
\begin{equation}
L = V\left(\frac{dI}{dt}\right)^{-1},
\end{equation}
with $V$ measured as the difference between the maximum voltage achieved during the ramp and the voltage response due to the applied current and constant resistance. Multiple ramp rates were used for an average inductance of $1.04$ mH, compared to the predicted $1.15$ mH. %$1.53$ mH. 

\section{Discussion}

Magnetic fields and total  resistance for P2 were measured at both $77$ K and $32-42$ K, with field measurements being in the expected ranges for the operating currents. Additional testing included the assembly, cooling, and cryogenic interfacing of additively manufactured multi-section magnets.%Additional testing included the viability of metal 3D-printing and additive manufacturing methods, the ability to assemble and cool magnets printed in multiple sections, and the interfacing of the magnet with conduction-based cryogenic systems while maintaining electrical insulation.

%The field of $4.5$ mT measured approximately $0.2$ m from axis at $110$ A agrees with calculations to within 90\%.
The magnetic field values agree closely with calculations. Further discrepancies may arise from slight changes in probe orientation, and the difficulty in defining a magnetic axis for a non-planar, asymmetric magnet. The resistance across the magnet is sufficiently low that operation at higher currents does not heat up the magnet itself significantly, though the copper leads showed significant heating (up to $20$ K increases) and better current lead thermal management is required. The magnet was quenched close to $120$ A due to the warming of the leads and the soldered HTS joints close to the leads. This highlights a potential weak-point of the design, and future designs will work to ensure adequate clamping and thermal management in this region.
%These HTS joints needed to be added to repair previous damage resulting from poor clamping. This highlights a potential weak-point of the design, and future designs will work to ensure adequate clamping and thermal management in this region.

%Further characterization of magnet parameters such as the $L/R$ time will impact the design process, particularly charging and discharging times. The inductance was calculated based on back EMF measurements, yielding an average inductance of $1.04$ mH. Using Grover’s thin-ring formula, the magnet’s inductance was estimated at $1.53$ mH, a difference of $\approx 32\%$. This discrepancy may arise from inaccurate model assumptions (thin-ring approximation, geometry, material properties, stray/lead inductance, etc.), experimental errors, or due to the current-dependent inductance observed in HTS magnets \cite{venuturumilli}.

Additional characterization of magnet parameters such as the $L/R$ time will impact the design process, particularly charging and discharging times. The inductance was calculated based on back EMF measurements, yielding an average inductance of $1.04$ mH, matching the prediction of $1.15$ mH to within $90\%$. The predicted value was calculated from the stored magnetic energy using a MLFMM magnetic field code. The discrepancy may lie in measurement uncertainties or effects such as radial current leakage \cite{deng, venuturumilli}. P3 can now be tested using the same setup, with higher fields allowing for higher fidelity measurements and more robust results. The upgrade to a conformal cryostat, as required by the full-scale coils, will enable greater flexibility regarding Hall probe measurements and diagnostics due to increased access. The use of parallel-wound HTS tapes for quench resilience will also be of interest.

%This discrepancy indicates that either the measurement setup or the model assumptions (thin-ring approximation, geometry, material properties, stray/lead inductance, etc.) are affecting the result; further investigation is required to reconcile the values.

%Next steps for both P2 and P3 include higher current and field tests, as well as further studying of the quench behavior. The tolerancing of 3D prints needs to be measured to ensure manufacturing errors are within quasi-symmetry error requirements. Finally, the experimental setup will be modified and used to test HTS joints at lower temperatures and at higher critical currents, to ensure sufficiently low resistance. The successful completion of these next steps and optimization of processes will allow for the design and fabrication of the full-scale CSX magnets.

\section{Conclusion}

This prototype campaign establishes the critical enabling technologies required to realize non-planar HTS coils for a compact, quasi-axisymmetric stellarator. Sequential design logic, beginning with the planar P1 magnet, followed by the high-strain P2 magnet, and ending in the concave, high-field P3 magnet, has efficiently isolated and reduced risk in manufacturing, winding mechanics, structural joining, cooling, and energizing. Additive manufacturing of multi-section aluminum frames with dovetail and bolted interfaces has proven mechanically robust while preserving geometric accuracy required for quasi-symmetry. The gimballed winding setup maintained perpendicular tape alignment and acceptable strain in non-planar channels, validating the strategy for more complex future coil shapes. Solder potting using SnPb solder has provided a passive pathway for current redistribution, directly addressing localized heating hazards associated with HTS quenches.

Experimental results corroborate design predictions: both P1 and P2 produced on-axis fields consistent with modeling at $77$ K; the cryogenic operation of P2 near $30–40$ K established effective conductive cooling through staged sapphire interfaces and demonstrated manageable thermal gradients under operating currents. Adjustments need to be made to reduce resistance and improve handling of heat through improved copper-HTS interfaces. Early lap joint tests have reached sub-$\mu \Omega$ resistance levels at $77$ K, supporting feasibility of the multi-kilometer tape lengths required by the $0.5$ T target field. The forthcoming characterization of joint performance at $20$ K under real-world conditions such as magnetic fields, mechanical strain, and higher transport currents will refine scaling laws for joint length versus resistance and impact final coil design.

%Remaining work focuses on (i) rigorous testing of P3 at high currents to approach the $0.5$ T objective, (ii) expanding diagnostic coverage (embedded Hall sensors, enhanced voltage tap arrays, RGA) for real-time quench, thermal, and vacuum profiling,  (iii) implementing conformal cryostat geometries to prototype full-scale coil cryostats, (iv) integrating and investigating co-wound HTS layers for improved quench protection, and (v) characterizing and tightening manufacturing tolerances relative to quasi-symmetry error allowances. 

Remaining work focuses on (i) rigorous testing of P3 at high currents to approach the $0.5$ T objective, (ii) expanding diagnostic coverage (embedded Hall sensors, enhanced voltage tap arrays, RGA) for real-time quench, thermal, and vacuum profiling, (iii) implementing conformal cryostat geometries that reproduce the spatial and thermal boundary conditions of the full CSX device, (iv) systematic characterization and optimization of the HTS–copper busbar interface, including contact resistance and thermal conduction pathways, and (v) tightening manufacturing tolerances relative to quasi-symmetry error allowances. Successful execution of these next steps will bring together the mechanical, electromagnetic, and thermal designs, enabling full-scale CSX coil construction and advancing the broader application of HTS technology to optimized stellarator configurations.

\section{Acknowledgements}
This work presents the first results on the experimental side of CSX; it would not have been possible without the work of, and countless meetings with, Elizabeth Paul, Antoine Baillod, and the entire CSX theory team. Thanks as well to many useful exchanges with Paul Huslage and the EPOS/APEX collaboration at the Max Planck Institute for Plasma Physics. The team is also grateful for the support of Columbia University internal funds and the Columbia Fusion Research Center. T.K. and A.H. gratefully acknowledge research funding support from the I.I. Rabi Scholars Program at Columbia University.

\vfill


\begin{thebibliography}{1}
\bibliographystyle{IEEEtran}

\bibitem{hartwig}
Z. S. Hartwig et al., ‘The SPARC Toroidal Field Model Coil Program’, IEEE Trans. Appl. Supercond., vol. 34, no. 2, pp. 1–16, Mar. 2024, doi: 10.1109/TASC.2023.3332613.

\bibitem{nash}
D. Nash et al., ‘Prototyping and Test of the “Canis” HTS Planar Coil Array for Stellarator Field Shaping’, IEEE Trans. Appl. Supercond., vol. 35, no. 7, pp. 1–14, Oct. 2025, doi: 10.1109/TASC.2025.3594533.

\bibitem{baillod}
A. Baillod, E. J. Paul, G. Rawlinson, M. Haque, S. W. Freiberger, and S. Thapa, ‘Integrating novel stellarator single-stage optimization algorithms to design the Columbia stellarator experiment’, Nucl. Fusion, vol. 65, no. 2, p. 026046, Feb. 2025, doi: 10.1088/1741-4326/ada6dd.

\bibitem{pedersen}
T. S. Pedersen et al., ‘Construction and Initial Operation of the Columbia Nonneutral Torus’, Fusion Science and Technology, vol. 50, no. 3, pp. 372–381, Oct. 2006, doi: 10.13182/fst06-a1258.

%\bibitem{pedersen2}
%. S. Pedersen et al., ‘The Columbia Nonneutral Torus: A New Experiment to Confine Nonneutral and Positron-Electron Plasmas in a Stellarator’, Fusion Science and Technology, vol. 46, no. 1, pp. 200–208, Jul. 2004, doi: 10.13182/fst04-a556.

\bibitem{riva}
N. Riva et al., ‘Development of the first non-planar REBCO stellarator coil using VIPER cable’, Supercond. Sci. Technol., vol. 36, no. 10, p. 105001, Oct. 2023, doi: 10.1088/1361-6668/aced9d.

\bibitem{huslage3}
P. Huslage et al., ‘A Non-Planar ReBCO Test Coil With 3D-Printed Aluminum Support Structure for the EPOS Stellarator’, IEEE Trans. Appl. Supercond., vol. 36, no. 3, pp. 1–5, May 2026, doi: 10.1109/TASC.2025.3640157.

\bibitem{paz-soldan}
C. Paz-Soldan, ‘Non-planar coil winding angle optimization for compatibility with non-insulated high-temperature superconducting magnets’, J. Plasma Phys., vol. 86, no. 5, Oct. 2020, doi: 10.1017/s0022377820001208.

\bibitem{huslage1}
P. Huslage et al., ‘Strain optimisation for ReBCO high-temperature superconducting stellarator coils in SIMSOPT’, J. Plasma Phys., vol. 91, no. 2, p. E71, Apr. 2025, doi: 10.1017/S0022377825000224.

\bibitem{huslage2}
P. Huslage, D. Kulla, J.-F. Lobsien, T. Schuler, and E. V. Stenson, ‘Winding angle optimization and testing of small-scale, non-planar, high-temperature superconducting stellarator coils’, Supercond. Sci. Technol., vol. 37, no. 8, p. 085010, Aug. 2024, doi: 10.1088/1361-6668/ad5382.

\bibitem{rogers}
J. S. Rogers, G. D. May, C. D. Coats, and P. M. McIntyre, ‘Dynamics of Current-Sharing Within a REBCO Tape-Stack Cable’, IEEE Trans. Appl. Supercond., vol. 33, no. 5, pp. 1–6, Aug. 2023, doi: 10.1109/TASC.2023.3245999.

\bibitem{kobayashi}
H. Kobayashi, Y. Nakada, H. Saotome, D. Miyagi, Y. Nagasaki, and M. Tsuda, ‘Numerical Evaluation on Current Behavior of No-Insulation Coils With Parallel HTS Tapes’, IEEE Trans. Appl. Supercond., vol. 33, no. 5, pp. 1–6, Aug. 2023, doi: 10.1109/TASC.2023.3251301.

\bibitem{xue}
S. Xue et al., ‘Current sharing in double-sided REBCO tapes’, Supercond. Sci. Technol., vol. 37, no. 7, p. 075006, Jul. 2024, doi: 10.1088/1361-6668/ad4e76.

\bibitem{lalitha}
S. L. Lalitha, ‘Low resistance splices for HTS devices and applications’, Cryogenics, vol. 86, pp. 7–16, Sep. 2017, doi: 10.1016/j.cryogenics.2017.06.003.

%\bibitem{radovinsky} A. Radovinsky, N. Martovetsky, and S. Kuznetsov, ‘Method of Quench Detection in HTS Magnets’, IEEE Trans. Appl. Supercond., vol. 35, no. 5, pp. 1–5, Aug. 2025, doi: 10.1109/tasc.2024.3520084.

%\bibitem{watterson}
%A. Watterson et al., ‘Performance Assessment of PIT VIPER Cables Following Long-Duration Solder Exposure During Manufacturing’, IEEE Trans. Appl. Supercond., vol. 35, no. 5, pp. 1–5, Aug. 2025, doi: 10.1109/tasc.2025.3527947.

\bibitem{chen} 
Y. Chen et al., ‘HTS Joint Resistance for High-Field Magnets: Experiment and Temperature-Dependent Modeling’, J Supercond Nov Magn, vol. 35, no. 5, pp. 1089–1098, May 2022, doi: 10.1007/s10948-022-06181-0.

%\bibitem{simsopt} M. Landreman, B. Medasani, F. Wechsung, A. Giuliani, R. Jorge, and C. Zhu, ‘SIMSOPT: A flexible framework for stellarator optimization’, JOSS, vol. 6, no. 65, p. 3525, Sep. 2021, doi: 10.21105/joss.03525.

%\bibitem{hu} C. Hu, Y. Lin, Y. Tan, L. Wang, and J. Geng, ‘A quench detection method for parallel co-wound HTS coils based on current redistribution’, Supercond. Sci. Technol., vol. 37, no. 2, p. 025007, Feb. 2024, doi: 10.1088/1361-6668/ad1a45.

\bibitem{venuturumilli} 
S. Venuturumilli, R. C. Mataira, R. W. Taylor, J. T. Gonzales, and C. W. Bumby, ‘Modeling HTS non-insulated coils: A comparison between finite-element and distributed network models’, AIP Advances, vol. 13, no. 3, p. 035317, Mar. 2023, doi: 10.1063/5.0135291.

\bibitem{deng} 
X. Deng et al., ‘An Experimental and Numerical Study on the Inductance Variation of HTS Magnets’, IEEE Trans. Appl. Supercond., vol. 25, no. 3, pp. 1–5, Jun. 2015, doi: 10.1109/TASC.2014.2377794.




\end{thebibliography}
\end{document}